\documentclass{article}
\usepackage{amssymb}
\begin{document}

\newcommand{\shalf}{\hbox{${\textstyle{\frac{1}{2}}}$}}
\newcommand{\M}{{\cal M}}
\newcommand{\T}{{\cal T}}
\newcommand{\C}{{\cal C}}
\newcommand{\Gal}{\mathtt{Gal}}
\newcommand{\Lor}{\mathtt{Lor}}
\newcommand{\IGal}{\mathtt{IGal}}
\newcommand{\ILor}{\mathtt{ILor}}
\newcommand{\Euc}{\mathtt{Euc}}
\newcommand{\SO}{\mathtt{SO}}
\newcommand{\OR}{\mathtt{O}}  
\newcommand{\Aut}{\mathtt{Aut}}
\newcommand{\Autc}{\mathtt{Aut}^{\rm c}}
\newcommand{\Aff}{\mathtt{Aff}}
\newcommand{\GL}{\mathtt{Gl}}
\newcommand{\reals}{\mathbb{R}}
\newcommand{\integers}{\mathbb{Z}} 
\newtheorem{proposition}{Proposition}
\newtheorem{theorem}{Theorem}
\newtheorem{requirement}{Requirement}

\title{Uniqueness of Simultaneity}
\author{Domenico Giulini                       \\
        Institute for Theoretical Physics      \\
        University of Z\"urich                 \\
        Winterthurerstrasse 190                \\
        CH-8057 Z\"urich, Switzerland               }
\date{November 14. 2000}

\maketitle

\begin{abstract}
We consider the problem of uniqueness of certain simultaneity 
structures in flat spacetime. \emph{Absolute simultaneity} is 
specified to be a non-trivial equivalence relation which is 
invariant under the automorphism group $\Aut$ of spacetime. 
$\Aut$ is taken to be the identity-component of either the 
inhomogeneous Galilei group or the inhomogeneous Lorentz group. 
Uniqueness of standard simultaneity in the first, and absence 
of any absolute simultaneity in the second case are demonstrated 
and related to certain group theoretic properties.  
\emph{Relative simultaneity} with respect to an additional 
structure $X$ on spacetime is specified to be a non-trivial 
equivalence relation which is invariant under the subgroup in 
$\Aut$ that stabilises $X$. Uniqueness of standard Einstein 
simultaneity is proven in the Lorentzian case when $X$ is an 
inertial frame. We end by discussing the relation to previous 
work of others.        
\end{abstract}

\section*{Introduction}
Simultaneity is a relational structure on or of spacetime which 
helps to globally organise events. It may or may not be thought
of as intrinsic property \emph{of} spacetime, depending on the
limitations on the amount of structure put on the set of events.
If it is definable solely by means of the structural elements 
assigned to spacetime one usually speak of \emph{absolute} 
simultaneity. In this case a natural question is whether it is unique. 
If it is not definable in such a way, one has to add some
further structure with the help of which we may then define some 
\emph{relative} simultaneity, `relative' to the added structure. 

Adding sufficiently much structure will always allow 
to define \emph{some} notion of relative simultaneity, even though 
such a definition will generally be far from unique. Uniqueness is 
no issue for practical applications, as many of the modern global 
navigational systems -- like GPS or LORAN-C -- demonstrate. For them 
to work it is sufficient that suitable and well defined methods for clock 
synchronisation exist. Typically, such methods will heavily depend
on the contingent properties of the physical situation at hand, that 
is, abstractly speaking, on the properties of specific solutions 
to the dynamical equations of many body systems (earth, satellites 
etc.). Such solutions will clearly not respect any of the 
fundamental spacetime symmetries of the underlying theory.

There is hence no question that abundant simultaneity structures 
can be defined \emph{on} spacetime. But this fact does not teach 
anything to the spacetime theorist. He or she is interested
in the structure \emph{of} spacetime and whether simultaneity may 
be regarded as (unique?) part of it. 
If spacetime does not provide a sufficiently rich structure by 
itself, the natural question would be `how much' structure 
is missing. That is, whether we can add some \emph{minimal} amount 
of structure with respect to which a (unique?) characterisation of 
simultaneity can be given. Here, `minimal' could, for example, 
mean that the added structure should be invariant under as many 
of the original symmetries as possible. This will be the intuitive 
idea behind the approach followed here. The added structure should of
course be physically interpretable and give rise to a physically 
`reasonable' definition of simultaneity. For example, one may 
regard any definition as inadmissible in which a physically realizable 
world-line is allowed to intersect a set of mutually simultaneous 
events in more than one point.  This will also be a criterion which 
we adopt in this paper.

Our discussion is close in spirit to that of Malament~(1977) and 
Sarkar \& Stachel~(1999), and was in fact motivated by them. 
These authors are also concerned with certain uniqueness properties 
of Einstein's definition of synchronisation. Their background is the
debate about the `conventionality-of-simultaneity-thesis', an 
updated summary of which was given by Janis~(1999). Here the 
issue of uniqueness comes in because one adopted strategy to refute 
this thesis is to first identify non-conventionality with uniqueness 
and then to prove the latter. Clearly, this identification 
can be challenged upon the basis that every proof of uniqueness 
rests upon some hypotheses which the simultaneity relation is 
supposed to satisfy and which may themselves be regarded as 
conventional. Arguments of this kind were brought forward in 
particular by Janis (1983) and Anderson et al. (1998, 123-126)
and merely point towards a certain ambiguity in the possible 
meaning and range of 
the word `convention'.\footnote{Janis (1983, 101)
characterises a simultaneity structure as free of conventions, 
if it is `singled out by facts about the physical universe'.
But this is potentially also full of ambiguities. For example, 
are solutions to equations of motion which are actually realised 
in nature considered as `facts about the physical universe'?
If yes, and this is hard to deny, we can use them to define 
arbitrarily many (possibly practically very useful) simultaneities. 
To choose between those clearly requires a convention. On the other 
hand, on the largest observable scales our world is well described 
by a `cosmological' solution in which the electromagnetic microwave 
background radiation singles out a preferred (irrotational) 
congruence of local observers whose orthogonal spatial 
hypersurfaces define a preferred simultaneity structure.
Does the `cosmological' character sufficiently distinguish this 
solution to let us conclude non-conventionality?}
For this reason we concentrate on the question of uniqueness, 
which seems to be a much better behaved notion about which
definite statements can be made once conditions for simultaneity 
relation are specified.

There are various difficulties with the uniqueness arguments 
by Malament~(1977) and Sarkar \& Stachel~(1999). Whereas the 
arguments given by Malament appear mathematically correct, 
we agree with Sarkar \& Stachel that he puts 
\emph{physically} unwarranted restrictions on the simultaneity 
relation.\footnote{For fairness one should say that 
Malament had a different motivation, namely to prove that the 
standard simultaneity relation of special relativity is 
uniquely definable in terms of the relation of causal 
connectibility. Since the latter is invariant under a 
strictly larger group than the group of physical symmetries 
of spacetime, he had indeed good reasons to put the stronger 
invariance requirement. Note that this makes existence less and 
uniqueness -- provided existence holds -- more likely.} 
On the other hand, the mathematical 
arguments given by Sarkar \& Stachel are mathematically 
incomplete. Our last section is devoted to a more detailed discussion 
of these issues. In this paper we wish to present a fresh and
systematic approach from first principles which, we believe, is 
free from the uncertainties just mentioned.

Let us point out right at the beginning that our requirements on
simultaneity differ slightly from the ones used by 
Malament and Sarkar \& Stachel: 
they require the simultaneity-defining equivalence relation to be 
invariant under all \emph{causal} automorphisms (explained in 
section~\ref{sec:others}), whereas we only require invariance 
under spacetime automorphisms (to be defined below), which form 
a proper (i.e. strictly smaller) subgroup of the former. 
This implies that a priori our invariance-requirement is weaker 
and will therefore generally allow for \emph{more} invariant equivalence
relations. Uniqueness results in our setting should therefore be
considered as stronger. But our actual motivation for sticking with
spacetime automorphisms is simply that those causal automorphisms which 
are not spacetime automorphisms, like constant scale transformations, are 
no \emph{physical} 
symmetries.\footnote{We argue on the level of modern
classical and quantum field theories in flat space, thereby 
\emph{ignoring} General Relativity.}
For this reason we will also not include space- and time-reflections 
in the group of spacetime automorphisms. The latter were also excluded 
by Sarkar \& Stachel but not by Malament.
Since our presentation aims to be 
self-contained, it will also contain a fair amount of background material.

\section{Flat Spacetime and its Automorphisms}
Throughout we deal with flat spacetime which we denote 
by $\M$. It consists of a manifold together with certain 
geometric structures. The manifold is assumed to be 
diffeomorphic to $\reals^4$ with its natural differentiable 
structure. We think of $\reals^4$ as being endowed with a 
basis which is fixed once and for all, unless explicitly 
stated otherwise. Linear transformations are then interpreted 
as diffeomorphisms (active point transformations), not as 
changes of bases.

An effective method to (implicitly) specify geometric structures 
is via the choice of an \emph{automorphism group} $\Aut$, which 
is a subgroup of the group of bijections of $\reals^4$. 
In most cases, like in ours, it will turn out to be a finite
dimensional Lie group which acts by diffeomorphisms, but a priori  
this need not necessarily be so. Once the choice of $\Aut$ is made, 
a geometric structure is said to exist on, or be a property of, 
spacetime $\M$ 
iff\footnote{Throughout we use `iff' as abbreviation 
for `if and only if'.} 
this structure is invariant under  
$\Aut$.\footnote{This is the central idea of the 
`Erlanger Programm' of Felix Klein (1893): to characterise 
a geometry (in a generalised sense) by its automorphism group 
(Klein calls it `Hauptgruppe'). Relations belong to that geometry 
(are `objective' according to Weyl (1949, Chap.~III.13)) iff they 
are invariant under $\Aut$. This statement stands independent of 
the logical question of whether any such `objective' relation is 
actually \emph{derivable} or \emph{definable} within a given 
axiomatic setting (Weyl 1949, 73). This issue has recently been 
raised again by Rynasiewicz~(2000) in the context of the 
`conventionality of simultaneity' debate.}
Invariant structures are sometimes called `absolute' (like absolute
simultaneity), but one should keep in mind that this notion of
\emph{absolute} depends on, and is hence relative to, the choice 
of $\Aut$. That choice should really be considered as a 
\emph{physical} 
input.\footnote{Eventually it depends on the 
fundamental dynamical laws of quantum field theory (without 
gravitation), which denies the notion of empty space even 
locally. What we call the automorphisms of spacetime is the 
stabiliser of the ground state (`vacuum') within the group of 
dynamical symmetries of the theory. Note that this point of view
eventually also denies that there exists a fundamental distinction 
between kinematical and dynamical symmetries.}

In this paper $\Aut$ is either the inhomogeneous Galilei or the
inhomogeneous Lorentz group. Let us briefly recall the
essential requirements which lead to these groups.
\begin{itemize}
\item
Elements of $\Aut$ should be \emph{bijections} of $\reals^4$;
that is, they should be maps which are injective (same as `into')
and surjective (same as `onto'). Hence each transformation has an
inverse and no point (event) `gets lost' in a transformation.
Note that a priori we do not require transformations to be continuous 
or even smooth, which -- if you think about it --  would be hard to
justify physically. In fact, smoothness will be implied by this and 
the next condition.
\item
We assume we are given `forceless point-particles', that is, 
elementary point objects whose inertial trajectories define an
affine structure on $\M$ with respect to which the trajectories become
`straight lines'.  $\Aut$ is now required to preserve this affine
structure, i.e., transformations in $\Aut$ must map straight lines to
straight lines.
\end{itemize}
It is the main result of real affine geometry that bijections of
$\reals^n$  ($n\geq 2$) which map straight lines to straight lines
must be affine maps: $x\mapsto Ax+a$, where $A$ is an invertible
$n\times n$-matrix and $a\in\reals^n$. A proof of this fact is 
given in sections 2.6.3-4 of 
(Berger 1987).\footnote{To complete Berger's (1987) argument one 
needs to supply a proof of his proposition 2.6.4, which states that 
there are no non-trivial automorphisms of the real numbers. This 
well known fact can be shown in an elementary fashion.}
Hence $\Aut$ must be a subgroup
of the real affine group in 4-dimensions, called $\Aff(4,\reals)$, which 
is given by the semi-direct product $\reals^4\rtimes \GL(4,\reals)$ 
of the group of translations ($\reals^4$) with the group of general
linear transformations ($\GL(4,\reals)$). For $(a',L')$ and $(a,L)$ 
in $\Aff(4,\reals)$ their multiplication law reads:
\begin{equation}
(a',L')(a,L)=(a'+L'a,L'L).
\label{affine-mult}
\end{equation}  
\begin{itemize}
\item
We assume that all spacetime-translations are part of $\Aut$ 
(`homogeneity of spacetime'). Hence $\Aut$ is of the form
$\reals^4\rtimes\Aut^*$, where $\Aut^*\subseteq\GL(4,\reals)$. 
Of $\Aut^*$ it is further assumed that it contains the spatial 
rotations (`isotropy of space') as matrices of the 
form:\footnote{Vectors in $\reals^3$ carry an 
arrow overhead and are considered as $1\times 3$-matrices. 
The superscript $\top$ denotes matrix-conjugation. 
$4\times 4$-matrices are written in 
$\hbox{time}\oplus\hbox{space}$ - form.}
%
\begin{equation}
R(D)=\pmatrix{1      & {\vec 0}^{\top}\cr
              \vec 0 & D              \cr}\,,
\label{rotations-in-aff}
\end{equation}
where $D\in\SO(3)$. Note that we did not include space- and
time-reflections.
\item
We assume the relativity principle to hold, which says that 
velocity transformations $B(\vec v)$ (called boosts) are part 
of $\Aut^*$. The boosts are assumed to be continuously and 
faithfully labelled by $\vec v\in V\subseteq\reals^3$, where 
$V$ is connected. Finally, let $R(D)$ be as in 
(\ref{rotations-in-aff}), then we assume the following equivariance 
condition\footnote{Condition (\ref{rotation-boost-equivariance})
is usually not stated explicitly, but tacitly assumed in statements 
to the effect that one may w.l.o.g. (sic) restrict attention to 
boosts in a preferred direction (Berzi \& Gorini 1969, 1519), and 
that rotations about the $\vec v$-axis necessarily (sic) 
commute with $B(\vec v)$ (Torretti 1996, 80). 
On the other hand, Berzi \& Gorini  and
Torretti explicitly make use of 
(\ref{rotation-boost-equivariance}) but with $R$ being a spatial 
reflection which reverses the boost direction 
(Torretti 1996, 79), which unnecessarily 
involves reflection transformations (which we exclude), whereas 
the same can be achieved by choosing for $R$ a $\pi$-rotation about 
an axis $\perp$ to the boost direction.},
which should be regarded as part of the requirement of `isotropy of 
space':
\begin{equation}
R(D)B(\vec v)[R(D)]^{-1}=B(D\vec v)\,.
\label{rotation-boost-equivariance}
\end{equation}
\end{itemize}

Given these conditions marked with $\bullet$, one can rigorously 
show that $\Aut$ is either the inhomogeneous Galilei or the 
inhomogeneous Lorentz group for 
some yet undetermined velocity parameter $c$. The identification 
of $c$ with the velocity of light is a logically independent step 
which need not concern us here. The idea to just use the relativity
principle and not the invariance of the velocity of light in order 
to arrive at (something close to) the Lorentz group was first 
consistently spelled out by Frank \& Rothe (1911). 
The way sketched here is mathematically more complete and partly 
based on the work of Berzi \& Gorini (1969). 

\section{Simultaneity}
Simultaneity, $S$, is a \emph{relation} on $\M$, that is, a subset of
$\M\times\M$. If $(p,q)$ belongs to this subset we write $S(p,q)$,
which stands for the statement: `the point (event) $p$ on $\M$
is in relation (later called `simultaneous') to the point $q$'. 
More precisely, we require $S$ to be an \emph{equivalence relation}, 
which means that it ought to satisfy the following three conditions:
\begin{eqnarray}
&&
S(p,p)\ \forall p\in\M
\qquad\hbox{(reflexivity)}\,,
\label{e-rel-reflexivity}\\
&&
S(p,q)\Rightarrow S(q,p)\ \forall p,q\in\M
\qquad\hbox{(symmetry)}\,,
\label{e-rel-symmetry}\\
&&
S(p,q)\ \hbox{and}\ S(q,r)\Rightarrow S(p,r)\
\forall p,q,r\in\M
\qquad\hbox{(transitivity)}\,.
\label{e-rel-transitivity}
\end{eqnarray}
It is hard to see how one could do without the first two 
conditions, but transitivity is certainly not needed 
in order to talk about the simultaneity of \emph{pairs} of 
events. For example, it already allows to synchronise each member
of a set of clocks with a preferred `master-clock', which is 
indeed sufficient for certain practical purposes. However, 
transitivity \emph{is} needed in order to consistently talk 
about \emph{mutually} simultaneous events in sets of more 
than two.  

\subsection{Equivalence Relations}
Let us recall a few general properties of equivalence relations 
which we will frequently use. An equivalence relation 
$S$ on a set $\M$ is the same thing as a partition of $\M$. 
Recall that a `partition' is defined to be a covering by non-empty,
mutually disjoint sets. In the present context such sets are called 
equivalence classes. The equivalence class in which $p$ lies is 
called $p$'s equivalence class or $[p]$, and given by  
\begin{equation}
[p]:=\{q\mid S(p,q)\}\,.
\label{def-e-class}
\end{equation}
This definition makes sense since $[p]$ and $[q]$ are either disjoint 
or identical. Before showing this, we first note that reflexivity 
implies $p\in[p]$. Hence no $[p]$ is empty and each $p$ lies in some 
equivalence class. Now, if $S(p,q)$ then $[p]=[q]$ since symmetry and 
transitivity immediately imply that $S(p,r)$ 
iff $S(q,r)$. Moreover, in the same way we see that 
$r\in [p]\cap[q]$ implies $S(p,q)$ and consequently $[p]=[q]$,
which proves the claim.  Conversely, a partition $\M=\bigcup_iU_i$  
defines an equivalence relation through $S(p,q)\Leftrightarrow$ $p$ 
and $q$ lie in the same $U_i$. The conditions of reflexivity, 
symmetry, and transitivity are easily checked. Hence we have shown 
that an equivalence relation on $\M$ is the same as a partition of 
$\M$.

Two particularly boring equivalence relations are: 
1)~$[p]=[q]\ \forall p,q\in\M$ (just one equivalence class), and 
2)~$[p]\not =[q]\ \forall p,q\in \M\ \hbox{where}\ p\not =q$ 
(each point is a different class). We call an equivalence relation
non-trivial if it is different from these two.

\subsection{Invariant Equivalence Relations}
Suppose $\M$ carries an action of a Group $G$: 
$(g,p)\mapsto g\cdot p$. We say that the equivalence relation $S$ 
is invariant under this action iff 
\begin{equation}
S(p,q)\Leftrightarrow S(g\cdot p,g \cdot q)\,,
\quad\forall g\in G\,,\forall p,q\in\M\,.
\label{e-rel-invariance}
\end{equation}
Expressed in terms of the equivalence classes (\ref{def-e-class}) 
this is the same 
as{}\footnote{For $U\subseteq\M$ or $H\subseteq G$ we write:
$g\cdot U:=\{g\cdot p\mid p\in U\}$ and 
$H\cdot p:=\{g\cdot p\mid g\in H\}$.}
%
\begin{equation}
[g\cdot p]=g\cdot [p]\,,\quad \forall g\in G\,,
                              \forall p\in\M\,.
\label{e-class-invariance}
\end{equation}
\emph{Proof.} That (\ref{e-rel-invariance}) implies 
(\ref{e-class-invariance}) is seen as follows:
\begin{eqnarray*}
[g\cdot p] &=& \{q\mid S(g\cdot p,q)\}             \\        
           &=& \{q\mid S(p,g^{-1}\cdot q)\}        \\   
           &=& \{g\cdot r\mid S(p,r)\}             \\       
           &=& g\cdot\{r\mid S(p,r)\}=g\cdot[p]\,. 
\end{eqnarray*}
Conversely, if $S(p,q)$ then $q\in[p]$ and 
(\ref{e-class-invariance}) implies $g\cdot q\in[g\cdot p]$ so 
that $S(g\cdot p,g\cdot q)$. This proves the equivalence of 
(\ref{e-rel-invariance}) and (\ref{e-class-invariance}).

As already mentioned in the introduction, we regard a 
$G=\Aut$--invariant equivalence relation as a physical property 
of spacetime. Our central requirement on `absolute simultaneity' 
then reads as follows: 
\begin{requirement}
\emph{Absolute Simultaneity} is a non-trivial $\Aut$-invariant 
equivalence relation on $\M$ each equivalence class of which 
intersects any physically realizable trajectory in at most one point.
\label{def:abs-sim}
\end{requirement}
If no such absolute structure exists, we need to add some further
structural elements $X$ to $\M$. $X$ could be a subset of $\M$, 
like a single wordline which models an individual \emph{observer}, as 
in Malament (1977), or a whole 3-dimensional family of such observers
which define a \emph{reference frame}, as in Sarkar \& Stachel (1999).
Straight lines or families of straight lines correspond to inertial observers  
and inertial reference frames respectively. Now, let $\Aut_X$  
be the subgroup of $\Aut$ that preserves (stabilises) $X$. For example,
if $X$ is a subset of $\M$, this means that $\Aut_X$ should map 
points of $X$ to points of $X$ (pointwise $X$ need not be fixed), 
and if $X$ is a partition of $\M$ by subsets, like a foliation by 
straight lines, it means that $\Aut_X$ should preserve this partition,
i.e., map lines to lines. A relation is then said to exist on $\M$
relative to $X$, or be a property of $(M,X)$, iff it is invariant 
under 
$\Aut_X$.\footnote{
Again this notion of relative existence of geometric
relations may be found in Klein's `Erlanger Programm' 
(Klein 1893, \S2).} 
In other words, the relation is required to break none of the 
residual spacetime symmetries which still exist relative to 
the structure $X$. Our central requirement on `relative simultaneity' 
then reads as follows:
\begin{requirement}
\emph{Simultaneity relative to $X$} is a non-trivial $\Aut_X$-invariant  
equivalence relation on $\M$ each equivalence class of which 
intersects any physically realizable trajectory in at most one point.
\label{def:rel-sim}
\end{requirement} 

We see that in order to classify simultaneity-structures we essentially 
need to classify $G$-invariant equivalence relations, where $G$ is 
$\Aut$ or some subgroup thereof. This will be done to some extent in the 
following sections. There we will make extensive use of the following 
simple observations: Let $G_p$ denote the stabiliser subgroup of 
$p\in\M$ in $G$, that is, $G_p:=\{g\in G\mid g\cdot p=p\}$. 
If $S(p,q)$ then  
(\ref{e-rel-invariance}) immediately implies $S(p,g\cdot q)$ for 
all $g\in G_p$. Hence the whole $G_p$-orbit of $q$, denoted by
$G_p\cdot q$, lies in  $[p]$. Moreover, suppose $S(p,q)$ and that 
for some $g$, with $p'=g\cdot p$ and $q'=g\cdot q$, we have 
\begin{equation}
G_p\cdot q\cap G_{p'}\cdot q'\not =\emptyset\,,
\label{orbits-intersect}
\end{equation}
then $[p]=[p']$. The proof is simple: since the orbits $G_p\cdot q$ 
and $G_{p'}\cdot q'$ lie in $[p]$ and $[p']$ respectively, 
$[p]$ and $[p']$ intersect and must hence be equal.

\subsubsection{Existence}
One may ask for general criteria for when $G$-invariant
equivalence relations exist. For example, assume $G$'s action  
on $\M$ to be 2-point-transitive, which means that for any set of 
four mutually distinct points $p_1,p_2,q_1,q_2$ there exists 
a $g\in G$ such that $g\cdot p_1=q_1$ and  $g\cdot p_2=q_2$.
This is equivalent to saying that the stabiliser subgroups $G_p$ 
act transitively. Then, obviously, the only invariant equivalence
relation is the trivial one where $[p]=\M$ for all $p$. 
On the other hand, if $G$'s action is not 
transitive but still non-trivial, we can, for example, 
just set $[p]:=G\cdot p$ to define a non-trivial
$G$-invariant equivalence relation. Hence the mathematically 
most interesting situations arise when $G$ acts transitively
but not 2-point-transitively. This is precisely the situation we 
are dealing with.  Due to the spacetime translations, $\Aut$ clearly  
acts transitively on $\M$, but the stabiliser subgroup of, say, 
the origin, $\Aut^*$, does not. Its orbits consist of 3-dimensional 
submanifolds which are planes in the Galilean and hyperbola or 
light-cones in the Lorentzian case. 

A general criterion for the existence of $G$-invariant 
equivalence relations, or equivalently, $G$-invariant partitions, 
does indeed exist. Before we state it, recall that a subgroup $K$ 
of $G$ is called `maximal' iff there is no proper subgroup $H$ of 
$G$ which properly contains $K$. We have
\begin{theorem} 
Let $G$ act transitively on $\M$. There exists a non-trivial 
$G$-invariant equivalence relation on (equivalently: partition of) 
$\M$ iff the stabiliser subgroups $G_p$ are not maximal. 
\label{theorem-1}
\end{theorem}
Note that maximality either applies to all or none of the stabiliser
subgroups, since for a transitively acting $G$ they are all conjugate:
$G_p=g\cdot G_q\cdot g^{-1}$ if $p=g\cdot q$.
A proof of Theorem~\ref{theorem-1} may be found as proof of  
Theorem~1.12 in Jacobson (1974). We will not make 
essential use of this theorem because we prefer to give direct 
arguments. But it is still useful to know since it highlights 
a group theoretic property  (maximality of stabiliser subgroups)
that distinguishes the inhomogeneous Galilei from the 
inhomogeneous Lorentz group and which pinpoints the mathematical 
origin of their different behaviour regarding the existence of 
absolute simultaneity-structures.

\section{Galilean Relativity}
We speak of \emph{Galilean relativity} if $\Aut$ is the inhomogeneous 
(including translations), proper (no space reflections), orthochronous 
(no time reflection) Galilei group, which we denote by $\IGal$. 
According to the general results given above we only need to specify 
its homogeneous part $\Aut^*$. It is given by the
homogeneous Galilei group $\Gal$, which is the semi-direct product of
spatial rotations ($R\in\SO(3)$) and boosts ($\vec v\in\reals^3$). 
Hence we have 
\begin{equation}
\IGal=\reals^4\rtimes(\reals^3\rtimes\SO(3))\,,
\label{structure-IGal}
\end{equation}
where the first $\rtimes$ on the right side comes from  
(\ref{affine-mult}) and the second corresponds similarly to the 
law $({\vec v}',R')(\vec v,R)=({\vec v}'+R'\vec v,R'R)$. It is 
implemented by letting $Aut^*\subset\GL(4,\reals)$ be the 
subgroup of $4\times 4$ -- matrices of the form:
\begin{equation}
\pmatrix{1      & {\vec 0}^{\top}\cr
         \vec v & R              \cr}\,.
\label{def-Gal}
\end{equation} 
Note that the first semi-direct product in (\ref{structure-IGal})
is such that the action of $\Aut^*$ on $\reals^4$ is not 
irreducible: it leaves invariant the subgroup $\reals^3$ 
of spatial translations (due to the zero-vector in the upper right
corner of (\ref{def-Gal})). Consequently, 
$\reals^3\rtimes\Aut^*$ is a subgroup of $\Aut$.
Note also that the same is not true for time translations. 
With respect to Theorem~\ref{theorem-1} this implies that $\Aut^*$ -- 
the stabiliser subgroup of the origin in $\reals^4\cong\M$ -- is 
not a maximal subgroup of $\Aut$, since we can still adjoin the 
group $\reals^3$ of spatial translations. Hence Theorem~1 
guarantees the existence of a non-trivial $\Aut$-invariant 
equivalence relation. But this will be proven directly below.

$\IGal$ is parameterised by ten real numbers: three for   
$R$, three for a boost-vector $\vec v$, three for a spatial 
translation vector $\vec a$ and one for a time-translation $b$. 
A general element $g\in\IGal$ can then be uniquely labelled  
by $(R,\vec v,\vec a,b)$. The law for multiplication and 
inversion then simply 
read\footnote{
$\IGal$ can be embedded in $\GL(5,\reals)$ as follows:
\begin{equation}
g(R,\vec v,\vec a,b)\longrightarrow 
\pmatrix{R & \vec v & \vec a\cr
         0 & 1      & b     \cr
         0 & 0      & 1     \cr}\quad\in\GL(5,\reals)
\label{IGal-matrix}
\end{equation}
which also gives (\ref{IGal-mult},\ref{IGal-inv}). In this picture 
$\M$ is identified with the 4-dimensional hyperplane $x^5=1$, 
which then leads to (\ref{IGal-action-x},\ref{IGal-action-t}).
}
\begin{eqnarray}
g'' &=&
     (R'R                          \,,\,
     {\vec v}'+R'\vec v            \,,\,
     {\vec a}'+R'\vec a +b'\vec v  \,,\,
     b'+b)                         \,,
\label{IGal-mult}
\\
g^{-1} &=&
        (R^{-1}                    \,,\,
         -R^{-1}\vec v             \,,\,
         -R^{-1}(\vec a-b\vec v)   \,,\,
         -b)                       \,.
\label{IGal-inv}
\end{eqnarray}    
Writing $p$ in $\M$ as $(t,\vec x)$, the action of 
$g$ on $p$ reads: 
\begin{eqnarray}
\vec x\mapsto {\vec x}' &=& R\vec x+\vec vt+\vec a\,,
\label{IGal-action-x}\\
t\mapsto t' &=& t+b\,.
\label{IGal-action-t}
\end{eqnarray}

The subgroup $\Euc\subset\IGal$ of Euclidean motions consists of 
spatial rotations and translations. It is given by all elements 
where $\vec v=\vec 0$ and $b=0$. 
Its orbits are the planes of constant $t$ which we denote by $\Sigma_t$. 
Note that boosts act like translations in each $\Sigma_t$ separately,
but scaled with a factor $t$. Only time-translations permute the
planes $\Sigma_t$. The general law is $g\cdot\Sigma_t=\Sigma_{t+b}$.

\subsection{Galilean Simultaneity
\label{galilean-simultaneity}}
Let $S$ be a $G=\IGal$--invariant, non-trivial equivalence relation. 
We choose a hyperplane $\Sigma_t$ and a point $p\in\Sigma_t$.

First we assume that a point $q\not = p$ exists on $\Sigma_t$ such
that $S(p,q)$. For the moment we forget about $\M$ and restrict
attention to $\Sigma_t$ which we regard as $\reals^3$ with standard
inner product and norm $\Vert\cdot\Vert$. $\Euc\subset\IGal$ acts
transitively on $\Sigma_t$ by standard Euclidean motions.
The stabiliser subgroup $\Euc_p$ of $p$ consists of all 
rotations centred at $p$. The orbit $\Euc_p\cdot q$ is a
2-sphere around $p$ of radius $\Vert p-q\Vert$.
Now, let $g\in\Euc$ be a translation by a vector of norm less then 
$2\Vert p-q\Vert$. Clearly $S(p',q')$ for $p'=g\cdot p$ and 
$q'=g\cdot q$.  Moreover, since the distance between $p$ and $p'$ is
less then $2\Vert p-q\Vert$, the $\Euc_p$-orbit of $q$ and the
$\Euc_{p'}$-orbit of $q'$ intersect, which implies $[p]=[p']$;
compare discussion surrounding (\ref{orbits-intersect}). 
Since any point on $\Sigma_t$ can be reached by a finite number of 
translations of norm less than $2\Vert p-q\Vert$, we only need to
iterate this argument to show that all points of $\Sigma_t$ lie 
in the same equivalence class. 

Next we assume that a point $p'\not\in\Sigma_t$ exists such that
$S(p,p')$. Let $p'$ be a member of, say,  $\Sigma_{t'}$, where
$t\not =t'$. Consider the stabiliser subgroup $\IGal_{p'}$
of $p'$. Besides certain rotations (which need not concern us at the 
moment), it contains the 3-dimensional subgroup which is given by 
the combinations of translations and boosts for which 
$\vec a=-t'\vec v$. By construction this subgroup fixes 
$\Sigma_{t'}$ pointwise, but acts transitively on $\Sigma_t$  
via translations of the form $\vec x\mapsto \vec x+(t-t')\vec v$. 
This already proves that $[p](=[p'])$ contains both hyperplanes, 
$\Sigma_t$ and $\Sigma_{t'}$. The latter is seen by just reversing 
the r\^oles of $p$ and $p'$ in the argument.

So far our arguments show that, for any $p\in\M$, $[p]$ is a union
of planes $\Sigma_t$ one of which contains $p$. It is consistent with
$\IGal$-invariance to choose for $[p]$ just the single plane 
containing $p$ since $g\cdot\Sigma_t=\Sigma_{t+b}$
implies (\ref{e-class-invariance}). We may thus call this the 
\emph{minimal} or \emph{finest} non-trivial $\IGal$-invariant 
equivalence relation on $\M$. If $[p]$ contains more than one 
plane, say $\Sigma_t$ and $\Sigma_{t'}$, then for $\IGal$ 
invariance (here only time-translations matter) it is necessary 
that all planes $\Sigma_{t+n(t'-t)}$ for $n\in\integers$ are also 
contained in $[p]$. Therefore, if $\lambda$ denotes the infimum of 
all time-differences of planes contained in $[p]$, then $[p]$ is 
the union of all planes $\Sigma_{t+n\lambda}$ for $n\in\integers$, 
where $p\in\Sigma_t$. If this infimum were zero we would obtain 
the trivial equivalence relation where all of $\M$ is a single 
class. Hence we have
\begin{theorem}
Let $S$ be a non-trivial $\Aut=\IGal$ -- invariant equivalence
relation on $\M$. Then the possible equivalence classes $[p]$ 
are given by: \\
(i)  the plane $\Sigma_t$ containing $p$;\\
(ii) the union over $n\in\integers$ of planes $\Sigma_{t+n\lambda}$, 
     where $0<\lambda\in\reals$ and $p\in\Sigma_t$. 
\label{th:IGal-inv-eq-rel}
\end{theorem}

Next to the mathematical space $\M$ that represents physical 
spacetime, one may also associate a mathematical space $\T$ 
that simply represents time, namely the set of equivalence 
classes given by the quotient 
\begin{equation}
\T:=\M/S\,.
\label{def-time}
\end{equation}
In case the equivalence classes in $\M$ just consist of single 
hypersurfaces $\Sigma_t$, $\T$ is isomorphic to $\reals$.
The action of $\IGal$ on $\T$ is just $(g,t)\mapsto t+b$.
In case  there are more than one $\Sigma_t$ in each 
equivalence class, $\T$ is isomorphic to the circle
$S^1=\reals/\{\hbox{identification mod $\lambda$}\}$. In this 
sense time is periodic with period $\lambda<\infty$. Note
however that this does not mean that we may make periodic 
identifications in $\M$ and represent spacetime by 
$\M_{\lambda}:=\M/\integers$, where $\integers\subset\IGal$ is
represented by the discrete 
time translations $t\mapsto t+n\lambda\,,n\in\integers$.
The point being that this space (homeomorphic to $S^1\times \reals^3$) 
would not support an action of $\IGal$, since the
boost transformations $(t,\vec x)\mapsto(t,\vec x+t\vec v)$ are 
incompatible with such an identification. This is due to the 
group-theoretic fact that time-translations do not form a 
\emph{normal} subgroup in $\IGal$, in contrast to spatial translations.
For example, periodic \emph{spatial} identifications by some 
integer lattice $\integers^3\subset\reals^3$, which certainly 
does form a normal subgroup, results in a closed spatial space whose 
topology is that of the 3-torus, $T^3$, and a spacetime 
$\M'=\reals\times T^3$ which still carries an action of $\IGal$, 
though not an effective one, since $\integers^3$-valued spatial
translations now act trivially. We summarise the results of this
section in
\begin{theorem}
Standard Galilean simultaneity is the unique absolute simultaneity 
satisfying Requirement~\ref{def:abs-sim} for $\Aut=\IGal$. 
It is also the unique non-trivial $\IGal$-invariant equivalence 
relation on $\M$ for which time is non-cyclic, or for which the 
equivalence classes in $\M$ are connected.   
\label{th:IGal-inv-sim}
\end{theorem}

\section{Lorentzian Relativity}
We now wish to explore the consequences of replacing $\IGal$ by 
the inhomogeneous Lorentz group $\ILor$ (also known as
Poincar\'e group). This group has not only quite a different 
group-structure than $\IGal$, but also very different orbits
in $\M$. This results essentially from the way 
boost-transformations are implemented, which are now not 
allowed to boost beyond a finite limit-velocity $c$, 
usually taken to be the velocity of light, but the value of $c$ is  
unimportant for us as long as $0<c<\infty$. In the following we 
choose units such that $c=1$.

$\ILor$ is obtained by choosing for $\Aut^*\subset\GL(4,\reals)$ 
the homogeneous (proper, orthochronous) Lorentz group, $\Lor$.
To define it, consider the real $4\times 4$-matrices $L$ which 
leave the diagonal-matrix -- called the Minkowski metric -- 
$\eta:=\hbox{diag}(1,-1,-1,-1)$ invariant in the following sense:
\begin{equation}
L\eta L^{\top}=\eta\,.
\label{def-Lor}
\end{equation}
Those $L$ with determinant $+1$ form the proper Lorentz group
which is also denoted by $SO(1,3)$, a notation with an obvious 
meaning in view of (\ref{def-Lor}). This group has two 
components: one where the time-time component $L_0^0\geq 1$, 
the other where it is $\leq -1$. The former is the identity 
component and leads to our group $\Lor$ of proper orthochronous 
Lorentz transformations. As for the Galilei group we excluded 
reflections of space or time. 

To see the group-theoretic difference between $\Gal$ and
$\Lor$, recall that $\Gal$ is a semi direct product
$\{\hbox{boosts}\}\rtimes\{\hbox{rotations}\}$, which implies 
that boosts and rotations separately form subgroups, and that the 
boost-subgroup is normal (i.e., invariant under conjugations in 
$\Gal$; compare 
(\ref{rotation-boost-equivariance})). $\Lor$, on the other hand, 
is a \emph{simple} group, that is, it does not contain any normal
subgroups other than the identity and the whole group. Rotations
still form a subgroup, but boosts do not. In general, a boost 
multiplied by a boost is a boost times a non-trivial rotation
(the latter being the origin of `Thomas-Precession').
Requirement (\ref{rotation-boost-equivariance}) still holds, 
of course, and merely means that boosts form an invariant 
\emph{set} under conjugation with rotations but not under all 
conjugations in $\Lor$.
  
The translation subgroup in $\ILor$ acts on $\M$ just in the 
same way the translations in $\IGal$ do. This is also true 
for spatial rotations, which together with boosts make up $\Lor$.
The former correspond to matrices of the form 
\begin{equation}
\pmatrix{1      & {\vec 0}^{\top} \cr
         \vec 0 & R               \cr}\,,
\label{rotations-in-Lor}
\end{equation}
where $R\in SO(3)$. On the other hand, boost with velocity
$\vec v=v\vec n$, where $v:=\Vert\vec v\Vert<1$, correspond to
matrices of the form ($\gamma:=1/\sqrt{1-v^2}$):
\begin{equation}
\pmatrix{\gamma        & \gamma{\vec v}^{\top} \cr
         \gamma\vec v  & \mathbf{1}+
         (\gamma-1)\vec n\otimes{\vec n}^{\top}\cr}\,,
\label{boost-in-Lor}
\end{equation}
which act like
\begin{eqnarray}
\vec x\mapsto{\vec x}' &=& 
\vec x +\gamma \vec vt+(\gamma-1)(\vec n\cdot\vec x)\vec n\,,
\label{lorentz-action-x}\\
t\mapsto t'            &=& 
\gamma(t+\vec v\cdot\vec x)\,.                                
\label{lorentz-action-t}
\end{eqnarray}
Whereas (\ref{lorentz-action-x}) is merely a deformation of 
the boost action in (\ref{IGal-action-x}), 
(\ref{lorentz-action-t}) differs significantly from 
(\ref{IGal-action-t}). Thinking of $\M$ as $\reals ^4$, it means 
that the family of parallel planes
$t=\hbox{const.}$, which can be characterised by their normal-direction 
(here w.r.t. standard Euclidean metric of $\reals^4$)
given by $(1,\vec 0)$, will be transformed into the family of 
tilted planes with normal direction given by $(1,\vec v)$, i.e.
tilted by the angle $\tan\alpha=v$. Hence \emph{any} two planes 
from the first and second family respectively intersect.

With respect to Theorem~\ref{theorem-1} we also remark that
in $\reals^4\rtimes\Aut^*$ the action of $\Aut^*=\Lor$ on 
$\reals^4$ is now irreducible, hence no subspace of translations 
is left invariant, as it was the case for spatial translations 
when $\Aut^*=\Gal$. This implies that $\Aut^*=\Lor$ and all its
conjugations by translations -- which make up the stabiliser 
subgroups $\Aut_p$ -- are maximal subgroups of $\Aut$. 
From Theorem~\ref{theorem-1} we can therefore anticipate that 
there will be no $\Aut=\ILor$-invariant non-trivial equivalence 
relation on $\M$. Below we prefer to give a simple direct proof 
of this fact.

\subsection{Lorentzian Simultaneity}
Since spacetime translations and spatial rotations in $\IGal$
and $\ILor$ act identically on $\M$, all the arguments which 
were given in the framework of Galilean relativity and which 
did not use boost do also apply in the present case. 
In particular, each equivalence classes of any non-trivial
$\ILor$-invariant equivalence relation $S$ contains one or more 
of the planes $\Sigma_t$. But now comes the point: since boosts 
transform the family of planes $\Sigma_t$ to a tilted family 
$\Sigma'_{t'}$, any member of which intersects any member of the 
former, all the planes must be in the same equivalence class. 
This implies
\begin{theorem}
The only $\ILor$-invariant equivalence relation on $\M$ is the
trivial one where $\M$ is the only equivalence class. 
Hence absolute simultaneity satisfying Requirement~\ref{def:abs-sim} 
does not exist in Lorentzian relativity.
\label{th:ILor-inv-sim}
\end{theorem}
The proof is almost trivial: since $\Sigma_t$ lies within a single 
class, and since boosts in $\ILor$ map classes to classes, the image
$\Sigma'_t$ of $\Sigma_t$ under a boost also lies within one class. 
But it intersects \emph{all} $\Sigma_t$ and hence all other 
classes, which implies that there is only one class. 

\subsection{Lorentzian Simultaneity Relative to an Inertial Frame
\label{sec:lor-sim}}
What we have just learned is that $\Aut^*=\SO(1,3)$ does not 
leave spacetime with enough structure to be able to define 
absolute simultaneity. But what about relative simultaneity? 
For this we have to add some further structure $X$.
Clearly, if we choose $X$ so that $\Aut$ gets broken down completely, 
$\Aut_X$-invariance will be an empty requirement and  \emph{any} 
equivalence relation will do. As outlined in the introduction, 
the task is to choose $X$ rich enough to ensure existence but 
otherwise as symmetry-preserving as possible. If this leads to 
a sufficiently big $\Aut_X$, the residual invariance
requirement may ensure uniqueness. 

The structure $X$ we wish to consider here is an \emph{inertial reference
frame}, which in our case (flat geometry) corresponds to a foliation of 
$\M$ by (necessarily parallel) timelike straight lines. Let now $X$ 
stand for such a foliation by lines which are all parallel to the
four-vector $v=(1,\vec v)$. The foliation $X$ is obviously invariant 
(meaning lines are transformed to lines) under all spacetime 
translations, and obviously not invariant under any non-trivial boost. 
In fact, we can w.l.o.g. assume $\vec v=\vec 0$, for otherwise let 
$B(\vec v)$ denote the boost which maps the $t$-axis to a line in $X$, 
which we call the $t'$-axis, and refer the whole situation to $t'$ and 
the planes $\Sigma'_{t'}$ perpendicular (w.r.t. Minkowski metric)
to the $t'$-axis. The full stabiliser group $\Aut_X=\ILor_X$ is now 
seen to consist of time translations and the group $\Euc$ of spatial 
Euclidean motions: $\Aut_X=\reals\times\Euc$. 

First, we can now use the spatial rotations in $\Euc$ to argue 
exactly as in Section~\ref{galilean-simultaneity}: if in
some $\Sigma_t$ there exist two different points for 
which $S(p,q)$, then the argument there shows that $[p]$ 
contains $\Sigma_t$. Since time translations  can map 
$\Sigma_t$ to any other hyperplane in this family, each 
hypersurface $\Sigma_t$ is contained in some equivalence class.  

Next suppose that for a $p\in\Sigma_t$ some $p'\in\Sigma_{t'}$ 
for $t'\not =t$ exists so that $S(p,p')$. Now we cannot proceed 
as in Section~\ref{galilean-simultaneity} since $\Aut_X$ contains 
no boosts. Instead we argue as follows: let $\ell_{p'}$ 
denote the straight line in $\M$ through $p'$ which is
parallel to the $t$-axis, and let $r$ be its point of intersection 
with  $\Sigma_{t}$. The stabiliser $\Euc_{p'}$ of $p'$ in $\Euc$ 
consists of rotations which in $\Sigma_{t'}$ rotate about 
$p'$ and in $\Sigma_{t}$ rotate about $r$. 

If $p\not =r$ we can move $p$ by an element of $\Euc_{p'}$ 
to get another point $q$ in $\Sigma_t$ for which $S(p,q)$. 
Hence we are back to the case above which now shows that 
$[p](=[p'])$ contains $\Sigma_{t}$ and $\Sigma_{t'}$, the 
latter again by reversing the r\^oles of $p$ and $p'$ in the 
argument.

This conclusion is avoided iff $r=p$, i.e., iff $p$ lies on 
$\ell_{p'}$. The conclusion that each equivalence class contains 
some hyperplanes is avoided iff \emph{any} two different 
$p,p'$ for which $S(p,p')$ lie on the same straight line parallel 
to the $t$-axis. There clearly are non-trivial $\Aut_X$-invariant
equivalence relations whose classes are contained in the straight 
lines parallel to the $t$-axis. These are readily classified:
$S(p,q)$ iff either $p,q$ are on the same such line, or  
$\tau^n_{\lambda}\cdot p=q$ for some $n\in\integers$, where 
$\tau_{\lambda}\in\Aut_X$ is the time translation 
$\tau\mapsto t+\lambda$.
This leads to the following classification of all possible 
equivalence relations:
\begin{theorem}
Let $X$ be a foliation of $\M$ by timelike straight lines and 
$S$ a non-trivial $\Aut_X=\ILor_X$ -- invariant 
equivalence relation on $\M$. Then the possible equivalence 
classes $[p]$ are given by:\\ 
  (i)~the plane $\Sigma_t\ni p$ perpendicular 
      (Minkowski metric) to the timelike 
      lines $X$;\\
 (ii)~the union over $n\in\integers$ of planes $\Sigma_{t+n\lambda}$, 
      where $0<\lambda\in\reals$ and $p\in\Sigma_t$;\\ 
(iii)~the line in $X$ through $p$;\\  
 (iv)~the union over $n\in\integers$ of points $\tau^n_{\lambda}\cdot p$,
      where $\tau_{\lambda}$ is the translation by an amount $\lambda>0$ 
      along the line in $X$ through $p$.  
\label{th:ILor_X-inv-eq-rel}
\end{theorem}

In case (ii) each straight timelike line intersects each equivalence
class a countably infinite number of times. In case (iii) each 
timelike line in $X$ is its own equivalence class and hence intersects 
an equivalence class in uncountably many points. In case (iv) the same is
true for countably many points. Therefore, these cases do not define
a notion of relative simultaneity satisfying Requirement~\ref{def:rel-sim}, 
since `physically realizable trajectories' will certainly include all
timelike straight lines
(inertial motion). On the other hand, case~(i) does 
satisfy the condition that each equivalence class is cut at most 
(in fact: exactly) in one point by each physically realizable
trajectory, which here, for definiteness, we may e.g. specify to 
be all timelike piecewise differentiable curves. Hence we have
\begin{theorem}
Einstein simultaneity is the unique relative simultaneity satisfying 
Requirement~\ref{def:rel-sim} for $\Aut_X=\ILor_X$, 
where $X$ denotes an inertial frame (=foliation of $\M$ by 
timelike straight lines).
\label{th:ILor_X-inv-sim}
\end{theorem}

\section{Relation to Work of Others
\label{sec:others}}
Here we shall basically focus on the work of Malament 
(1977) and Sarkar \& Stachel~(1999). 
Both are directly concerned with uniqueness issues, but neither gives a 
systematic classification of invariant equivalence relations.
Malament proves uniqueness of Lorentzian simultaneity relative 
to a single observer. In this case $X$ is a single timelike
straight line. But instead of spacetime automorphisms $\Aut$ he takes 
all \emph{causal} automorphisms, which we denote by $\Autc$, by 
which he understands all bijections $f$ of spacetime such that 
$p-q$ is non spacelike iff $f(p)-f(q)$ is non spacelike. 
It has been proven by Alexandrov (1975)
that any such transformation is a combination of transformations 
in $\ILor$, time reflections $(t,\vec x)\mapsto (-t,\vec x)$, 
space reflections $(t,\vec x)\mapsto (t,-\vec c)$, and dilatations
$p\mapsto\lambda p$ with 
$\lambda\in\reals_+$ (positive real numbers).\footnote{
The same is true for any bijection which in both directions 
preserves just `lightlike' or just `timelike' 
separations.  Moreover, in the time oriented case, the same 
statements hold if one restricts to just future (or past) 
oriented separations and if time reflections are eliminated from
the list of possible transformations. Note that, mathematically 
speaking, the particular non-trivial aspect of these results lies 
in the lack of any initial continuity requirement for the bijective 
maps; the listed requirements suffice to imply continuity.}
More precisely, Malament proved the following
\begin{theorem}[Malament 1977]
Let $X$ be an initial observer, i.e., a timelike straight line. 
Let $S$ be an $\Autc_X$-invariant non-trivial equivalence relation,
which also satisfies the following condition: there exists a point 
$p\in X$ and a point $q\not\in X$ such that $S(p,q)$. Then $S$ is 
given by standard Einstein simultaneity.
\label{th:malament}
\end{theorem}
In a recent review, Anderson et al. (1998, 124-125) 
claim Malament's proof to be technically incorrect.
We disagree, as do  Sarkar \& Stachel (1999) and apparently 
also Janis (1999). However, it is true that the proof presented by
Malament (1977) leaves out some details. To settle this technical issue, 
we present an alternative and somewhat more detailed proof in the 
Appendix which uses the language developed in previous sections. 

Let us look at the same situation from our point of view, where 
instead of $\Autc$ we take $\Aut=\ILor$. We may w.l.o.g. take
$X$ to be the time axis; otherwise we boost and translate the 
selected observer to rest at the origin and take the conjugate 
of all subgroups to be mentioned by that combination of a boost and a
translation. Then $\Aut_X=\reals\times\SO(3)$,  
where $\reals$ consist of pure time translations and $\SO(3)$ are 
the rotations in $\Euc$ fixing the $t$-axis. Picking a single inertial
observer out of an inertial reference frame eliminates the space
translations in $\Euc$. This distinguishes the present case from that 
discussed above and makes a big difference concerning the question 
of uniqueness. Consider the two-parameter family of subsets of $\M$:
\begin{equation}
\sigma(\tau,r):=
\{(t,\vec x)\in\M\mid t=\tau,\,\Vert\vec x\Vert=r\}, 
\label{def-sigma}
\end{equation}
given by the points (for $r=0$) of the $t$-axis and all 
concentric 2-spheres about the spatial origin in each $t=\hbox{const.}$ 
plane. This is already an $\Aut_X$-invariant partition of $\M$ so that 
taking the $\sigma(\tau,r)$ as equivalence classes would define a notion 
(though not a very reasonable one) of relative simultaneity satisfying
Requirement~\ref{def:rel-sim}. However, it obviously 
violates the condition in Malament's theorem, since no point on $X$ 
is related to a point off $X$. But this can be easily cured: 
just take $\Aut_X$-invariant unions of sets $\sigma(\tau,r)$  
which connect points on $X$ with points off $X$, and define these unions   
as new equivalence classes. For example, in each $t=\hbox{const.}$ 
hypersurface, we can take a central ball and spherical shells of 
radius 1: 
\begin{equation}
\sigma'(\tau,n):=
\bigcup_{r\in[n-1,n)}\sigma(\tau,r)\,,
\label{def-sigma-1}
\end{equation}
where $n$ is a positive integer. Another possibility would be
to unite sets 
$\sigma(\tau,r)$ in different hyperplanes $t=\hbox{const.}$ We just 
have to take care that no two sets which are causally related are in 
the same equivalence class. For example, as slight modification of 
(\ref{def-sigma-1}), we may take: 
\begin{equation}
\sigma''(\tau,n):=
\bigcup_{r\in[n-1,n)}\sigma(\tau+mr,r)\,,
\label{def-sigma-2}
\end{equation}
which also consist of an inner ball and concentric spherical shells,
but now taken from the half-cones $C^{\pm}(\tau,\alpha)$ with
vertex on $X$ at time $\tau$ and opening angle 
$\alpha=\vert\tan^{-1}(1/m)\vert$. They open to the future ($+$ sign) 
for $m>0$ and to the past ($-$ sign) for $m<0$. For the half-cones 
to be acausal, i.e. spacelike hypersurfaces, we need opening angles 
bigger than $\pi/2$, i.e., $\vert m\vert <1$. $m=1$ gives
future light-cones, $m=-1$ past light-cones.

This abundance of possibilities does not violate Malament's theorem, 
since the onion-like partitions of the spatial hypersurfaces 
$\Sigma_{\tau}$ or $C(\tau,\alpha)$ is not invariant under scale 
transformations; only the partition of $\M$ into the $\Sigma_{\tau}$ 
or $C(\tau,\alpha)$ ($\alpha$ fixed) is. But, in turn, the latter is 
not invariant under time reflections. This is why Malament's
theorem 
works. (See the appendix for more details.) 

Sarkar \& Stachel pointed out that time reflections had to be 
included in order to prove Malament's theorem, and that without 
them one could still have lightlike half-cones as equivalence 
classes (half-cones of other opening angles are also possible, 
or course). They assert -- and we agree with this -- that there
is no physical reason to require invariance under time 
reflections. But likewise is there no physical reason to include 
dilatations, which they do include in their definitions, although in
footnote~11 of their paper Sarkar \& Stachel (1999) explicitly state that
their `considerations are independent of a requirement of invariance 
under scale transformations' (i.e. dilatations). They do not mention 
that dropping dilatations adds a plethora of new invariant equivalence
relations, like those in (\ref{def-sigma-1})(\ref{def-sigma-2}) 
(where e.g. the shell thickness is totally arbitrary and may, in
addition, depend on $r$). 

Finally Sarkar \& Stachel (1999) consider the case where $X$ is an 
inertial frame and try to show uniqueness of standard Einstein
simultaneity. Expressed in our terminology they argue as follows: 
if $X$ is an inertial frame, $\Aut_X$ contains spatial translations
perpendicular (w.r.t. Minkowski metric) to the lines in $X$. The orbit 
of any point $p$ under these translations is the plane $\Sigma_t$
containing $p$. Since `they [the spatial translations] are not to 
affect the simultaneity relation, these translations 
\underline{must take each simultaneity hypersurface to itself}' 
(Sarkar \& Stachel 1999, 217), which are therefore given by the
$\Sigma_t$'s. But this does not provide a proof, since the underlined part 
does not follow. It is not true that the equivalence classes 
of a $G$-invariant equivalence relation must separately be 
$G$-invariant sets; only the \emph{partition} must be $G$-invariant,
but $G$ may well permute the equivalence classes. The precise
statement is given in equation~(\ref{e-class-invariance}). 
From our Theorem~\ref{th:ILor_X-inv-eq-rel} it is also clear that 
classifying invariant equivalence classes is not quite sufficient, one 
also needs to impose some condition which eliminates those relations 
whose classes contain timelike related points. No such condition is
mentioned by Sarkar \& Stachel (1999).

\section*{Appendix: Proof of Theorem~\ref{th:malament}}

By hypothesis there exist $p\in X$ and $q\not\in X$ such that
$S(p,q)$. W.l.o.g. 
we may take $X$ to be the $t$-axis and $p$ to be the origin;
otherwise we can boost the observer to rest and translate $p$
to the origin, and then consider the conjugates of all subgroups 
to be mentioned by that combination of a boost and a translation.
We set $q=(t',{\vec x}')$.

From Alexandrov's (1975) results we know that $\Autc$ is as
specified above and can hence infer that the stabiliser subgroup 
$G:=\Autc_X$ consist of the following transformations and 
combinations thereof:
1)~time-translations, 
2)~time reflections about any moment in time,
3)~the group $\OR(3)$ of all orthogonal spatial transformations 
   (i.e. including reflections), and
4)~all dilatations about points on $X$.
Consider the subgroup $G_p\subset G$ that fixes $p$ (the origin). 
It consists 
of (and combinations thereof):
2')~the single time reflections about time zero: $t\mapsto -t$,
3')~all of $\OR(3)$, and
4')~dilatations about $p$: $r\mapsto\lambda r\,,
    \forall r\in\M$ and $\lambda>0$.

We know that $G_p\cdot q\subseteq [p]$, hence we are interested 
in the $G_p$-orbit of $q$. The orbit of $q$ under transformations 
4'), including the point $p$, consists of the half-line
$L:\lambda\mapsto\lambda q$, $\lambda\geq 0$. The shape of the orbit 
of this half-line under all remaining transformations in $G_p$ 
crucially depends on whether $t'\not =0$ or $t'=0$ (i.e. whether or not
$p$ and $q$ lie in the same hyperplane perpendicular to $X$). 

\medskip
\noindent
\emph{Case 1:} $t'\not =0$. The angle between $X$ and the half line 
$L$ ending on $X$ is $\alpha=\tan^{-1}(\Vert {\vec x}'\Vert/\vert
t'\vert)$ with $0<\alpha<\pi/2$. Hence the orbit of $L$ under all
transformations in $\OR(3)$ -- which is the same as the
orbit under $\SO(3)$, so spatial reflections do not add
anything new -- is a half-cone with vertex $p$ and opening angle
$\alpha$, which opens to the future if $t'>0$ and to the
past if $t'<0$. Acting with the remaining time reflection 2') 
results in a (full) cone with same vertex and opening angle,
which we call $C(p,\alpha)$.

\medskip
\noindent
\emph{Case 2:} $t=0$. Now $L$ is perpendicular to $X$ ($\alpha=\pi/2$).
The orbit of $L$ under $\OR(3)$ is the plane $\Sigma_{t=0}$. 
The main difference to Case~1 is that now the time reflection 2') adds 
nothing new since it leaves $\Sigma_0$ (pointwise) fixed. Hence the full 
$G_p$-orbit is still $\Sigma_0$.
\medskip

Consider Case~1; then $[p]\supseteq C(p,\alpha)$ and hence by
(\ref{e-class-invariance}) 
$[g\cdot p]\supseteq g\cdot C(p,\alpha)=C(g\cdot p,\alpha)$ for 
all $g\in\Autc$. (The last equality is obvious if one writes 
$g$ as a linear transformation in $\Autc$, which leaves the 
light cone at the origin invariant, followed by a translation. 
But any $g$ can be so written since the translations form a normal 
subgroup.)
Taking all spacetime translations for $g$ shows
$[p]\supseteq C(p,\alpha)$ for all $p\in\M$. But for $0<\alpha<\pi/2$ the
cones $C(p,\alpha)$ and $C(p',\alpha)$ for any two $p,p'$ necessarily
intersect. This is indeed easy to see and needs not be proven here. 
Hence all equivalence classes intersect and $S$ is trivial.

Finally consider Case~2; then $[p]$ contains the hypersurface 
$\Sigma_0$. If it contains any other point we are back to case~1 
and $S$ is trivial. Hence a non-trivial $S$ would have $[p]=\Sigma_0$.
Using time translations in (\ref{e-class-invariance}) then shows that 
$[p]$ would likewise be given by the hyperplane perpendicular
to $X$ containing $p$. But this proves the theorem since the partition 
of $\M$ into the hyperplanes $\Sigma_t$ is indeed $\Autc$-invariant and
hence defines a non-trivial equivalence relation. 

\goodbreak

\end{document}